\newcommand{\hmpc}{\mbox{ } h^{-1} \mbox{ Mpc}}
\newcommand{\hkpc}{\mbox{ } h^{-1} \mbox{ kpc}}
\newcommand{\hmsol}{\mbox{ } h^{-1}  M_{\odot}}
\newcommand{\lcdm}{$\Lambda$CDM }

\documentclass[floatfix]{emulateapj}
\usepackage{epsfig,subfigure}

\bibpunct[; ]{(}{)}{;}{a}{}{,}

\begin{document}
\title{Detecting dark matter-dark energy coupling with the halo mass function}

\author{P.~M.~Sutter} \email{psutter2@uiuc.edu}
\affil{Department of Physics,
	     University of Illinois at Urbana-Champaign,
            Urbana, IL 61801-3080}

\and

\author{P.~M.~Ricker} \email{pmricker@uiuc.edu}
\affil{Department of Astronomy,
       University of Illinois at Urbana-Champaign,
             Urbana, IL 61801\\
		National Center for Supercomputing Applications,
      University of Illinois at Urbana-Champaign,
            Urbana, IL 61801}
            
\begin{abstract}
We use high-resolution simulations of large-scale structure 
formation to analyze the effects of interacting dark matter and 
dark energy on the evolution of the halo mass function.
Using a chi-square likelihood analysis, we find significant 
differences in the mass function 
between models of coupled dark matter-dark energy and standard 
concordance cosmology (\lcdm) out to redshift $z=1.5$. We also 
find a preliminary indication that the Dark Energy 
Survey should be able to distinguish these models 
from \lcdm
within its mass and 
redshift contraints. While we can distinguish the effects 
of these models from \lcdm cosmologies with different 
fundamental parameters, DES will require independent 
measurements of $\sigma_8$ to confirm these effects.
\end{abstract}
\maketitle

\section{Introduction}
\label{sec:introduction}
Multiple independent lines of evidence, including 
observations of the large-scale matter distribution~\citep[eg.][]{Percival},
cosmic microwave background fluctuations~\citep[eg.][]{WMAP5}, 
and type Ia supernovae~\citep{Perlmutter,Riess},
 suggest that our universe 
is dominated by two components: dark matter, which is probably 
a form of nonbaryonic matter, and dark energy, which is a name 
for the presently unknown cause of the 
observed acceleration of expansion. However, we still lack 
an understanding of any possible interactions between these two 
principal constituents~\citep{Bean}. In a previous paper 
\citep[][hereafter SR08]{Sutter}, we examined 
the role that interacting dark matter and dark energy would play 
in the development of one-dimensional Zel'dovich pancakes, an 
important idealized case useful for understanding structure formation. 
Here we extend that 
preliminary work to a more realistic three-dimensional 
simulation of the growth of dark matter halos
in an attempt to find ways to distinguish these 
models from standard cosmology.
 
In this paper, we study the effects of a Yukawa interaction 
between a single family of nonrelativistic dark matter (DM) 
particles and a scalar field 
that is responsible for the dark energy (DE). We follow closely the formalism 
developed by~\cite*{Farrar}. Such an interaction is initially attractive 
because it is motivated by particle physics~\citep{Amendola2}  and 
might provide a way to explain the apparent emptiness of 
the voids, as demonstrated numerically by~\cite*{Nusser}.

There has been considerable interest recently in studying the 
effects of these interactions on structure, both using 
an analytic approach~\citep{Mainini} and using 
direct simulations~\citep[eg.][and others]{Maccio, Manera}. 
However, the current numerical 
studies suffer from poor resolution, and these results cannot 
reliably be compared to simulations of standard cosmological 
structure formation. In this paper, we use high 
spatial resolution and careful analysis 
to accurately capture many dark matter halos 
for use in comparison.
 
The following is a brief 
summary of the equations we solve and our numerical techniques. 
In Section~\ref{sec:massFunction} we discuss modifications to the halo 
mass function. We use these mass functions to distinguish 
interacting DM-DE from standard concordance cosmology 
using a $\chi^2$ likelihood test. 
Additionally, we discuss the feasibility of using the Dark Energy 
Survey~\citep{Annis} to detect this coupling within its mass 
and redshift contraints. Finally, we determine the 
extent to which two specific models of interacting DM-DE 
can be differentiated from each other.

\subsection{Analytical Methods}
Compared to simulations of the full non-linear theory, we found 
in SR08 that the perturbation theory presented by 
Farrar and Peebles is very accurate in determining the evolution of 
structure, and hence we will maintain the perturbative approach 
and assume fluctuations in the scalar field are small.
Under perturbation theory, the homogenous part of the 
dark energy scalar field, $\phi_b$, evolves as
\begin{equation}
\label{eq:phiEoM}
	\ddot{\phi_b} + 3 \frac{\dot{a}}{a} \dot{\phi_b} 
	  + \frac{d V}{d \phi_b} + 3 \frac{\Omega_{m0} H_0^2}{8 \pi G}
	    \frac{1}{\phi_b} a^{-3}= 0,
\end{equation}
where $\Omega_{m0}$ is the dark matter 
particle fraction of the critical density  
and $H_0$ is the Hubble constant.  
A subscript of $0$ denotes the present-day value.
The dark matter particle equation of motion is
\begin{equation}
\label{eq:dmEoM}
\dot{\bf{v}}  + \left( 2 \frac{\dot{a}}{a} + \frac{\dot{\phi_b}}{\phi_b} \right)
 \bf{v} = - \left( 1 + \frac{1}{4 \pi G} \frac{1}{\phi_b^2} \right) 
   \nabla \Phi.
\end{equation}
Here $\Phi$ is the normal comoving gravitational potential, 
$a$ is the scale factor, $\bf{v}$ is 
the comoving particle peculiar velocity, 
and $\bf{x}$ is the comoving position. Throughout, dots 
refer to derivatives with respect to proper time $t$. 
Perturbations in the scalar field give rise to the fifth force 
on the right-hand side in the equation above.

The comoving potential satisfies the Poisson equation:
\begin{equation}
\label{eq:poisson}
 \nabla^2 \Phi = \frac{4 \pi G}{a^3} \left( \rho - \overline{\rho} \right),
\end{equation}
with $\rho$ as the comoving DM particle density. 
Here and throughout, an overline indicates a spatial average.

At the present epoch, the field 
behaves as a cosmological constant, so the potential term 
in Eq.(\ref{eq:phiEoM} dominates and has a value 
\begin{equation}
	V(\phi_{b,0}) = \Omega_{\Lambda 0} \rho_{crit}. 
\label{eq:potentialToday}
\end{equation}
At early times, the coupling to matter dominates the 
scalar field equation of motion, and Eq.~(\ref{eq:phiEoM}) reduces to
\begin{equation}
\label{eq:initialPhiDot}
	\frac{d \phi_b} {d t} = - \frac{H_0^2}{G} 
	\frac{3 \Omega_{m0} }{8 \pi \phi_0} \frac{1}{a^3} t,
\end{equation}
which we use to set the initial condition for $\dot \phi_b$. 

The DM particle also has a field-dependent mass
\begin{equation}
\label{eq:mass}
   m_{DM} = {m_{DM,0}} \frac{\phi_b}{\phi_{b,0}}.
\end{equation}
The Friedmann equation, neglecting radiation, curvature, and 
baryonic terms, becomes 
\begin{equation}
\label{eq:friedmann}
	\left( \frac{\dot{a}}{a} \right)^2 = 
	H_0^2 \Omega_{m0} \frac{\phi_b}{\phi_{b,0}} a^{-3} + 
	\frac{8 \pi G}{3} \left[ \frac{1}{2} \left( \frac{d \phi_b}{d t} 
	\right)^2 + V(\phi_b) \right].
\end{equation}

To simulate a comparative $\Lambda$CDM cosmology, 
we fixed $\phi_b$ to the value in Eq.(\ref{eq:potentialToday}) 
and prevented any interactions between the field and particles.

\subsection{Numerical Methods}
We study a general power-law potential:
\begin{equation}
\label{eq:powerLaw}
	V(\phi) = K / {\phi}^{\alpha},
\end{equation}
where we are free to choose the constants $K$ and $\alpha$. Based 
on the comments made by Farrar and Peebles and our analysis 
in SR08, we chose two 
combinations of parameters. 
These were selected for behaviors that were significantly different 
from standard cosmologies, but not drastic enough to rule them out 
with current observational contraints.
Table~\ref{tab:FandP} lists the parameter values, the guessed 
initial field value at our simulation initial redshift, 
and the field value today as calculated 
from Eq.~(\ref{eq:phiEoM}). 

\begin{table}
  \centering
	\begin{tabular}{|c|c|c|c|c|} 
		\hline
	Label & $ \alpha $ & $K (G^{1+\alpha/2} / H_0^2)$ 
	      & $\phi_{\mbox{init}} (G^{1/2})$ & $\phi_0 (G^{1/2})$\\
	\hline
	A & $-2$ & $0.03$   & $1.89$   & $1.72$  \\
	B & $6$  & $2.0$    & $1.80$  & $1.68$  \\
	\hline
	\end{tabular}
	\caption{Simulation potential parameter value choices.}
	\label{tab:FandP}
\end{table}

For our simulations we chose FLASH v2.5, an adaptive-mesh refinement (AMR) 
code for astrophysics and cosmology~\citep{Fryxell}. FLASH solves the 
N-body potential 
problem with a particle-mesh multigrid FFT method~\citep{Ricker}.
FLASH uses cloud-in-cell mapping for interpolating between 
the mesh and particles~\citep{Hockney} and 
a second-order leapfrog integration scheme 
for variable-timestep particle advancement. 
We modified the standard FLASH code by adding 
the additional drag and force terms 
from Eq.~(\ref{eq:dmEoM}). 
At each timestep, the particle mass is updated according to 
Eq.~(\ref{eq:mass}). We calculate the scale factor and scalar field 
value in-code by numerically solving Eqs.~(\ref{eq:friedmann}) 
and (\ref{eq:phiEoM}), respectively.
For a more detailed explanation of solving the scalar field 
equation , see SR08.

For all calculations, 
we used concordance parameter values of
 $\Omega_{m0} = 0.26$, $\Omega_{\Lambda 0}=0.74$, and 
 $H_0 = 100 h = 71 \mbox{ km s}^{-1} \mbox{ Mpc}^{-1}$. 
 All runs took place in a three-dimensional box 
measuring $128 \hmpc$ per
side with $256^3$ particles. For each model, including 
a $\Lambda$CDM reference, we performed 10 simulations 
with $512$ zones per side and an additional 4 
simulations with $1024$ zones per side to study lower mass ranges. 
There was no refinement of grid spacing. 
All simulations used the same initial conditions: 
unperturbed particle positions were situated on a grid, 
and the initial velocities and positions were perturbed 
using Guassian fluctuations normalized to 
$\sigma_8 = 0.751$. We assumed P(k) from a \lcdm cosmology. 
We used the GRAFIC2 code~\citep{Bertschinger}
to generate these initial conditions. 
All computations started at a redshift of $z=56.8$.

We used a friends-of-friends (FOF) routine to find halos. This algorithm 
builds lists of all particles that are within a certain maximum 
distance of their neighbors. For all results, we chose a linking 
distance of $1/5$ of the unperturbed particle spacing, which is $500 \hkpc$. 
At every analysis redshift, we calculated the minimum resolvable halo 
particle count according to the prescription in~\citet{Lukic}:
\begin{equation}
n_{h,min} = \frac{\Delta (1.62 n_p/n_g)^3}{\Omega_{m0} (1+z)^3 } 
\left[ \Omega_{m0}(1+z)^3 + \Omega_{\Lambda 0} \right],
\label{eq:minMass}
\end{equation}
where $n_g$ and $n_p$ are the number of zones and particles 
per side, respectively. We chose an overdensity factor of $\Delta=200$.
To examine the halo mass function, we corrected the halo FOF 
particle counts
by the factor given in~\citet{Warren}:
\begin{equation}
{n_h}^{corr} = n_h (1-{n_h}^{-0.6}).
\label{eq:haloCorrection}
\end{equation}

\section{The Halo Mass Function}
\label{sec:massFunction}
\begin{figure*}
	\centering
  	\subfigure[Model A - $\Lambda$CDM]
		{\epsfig{figure=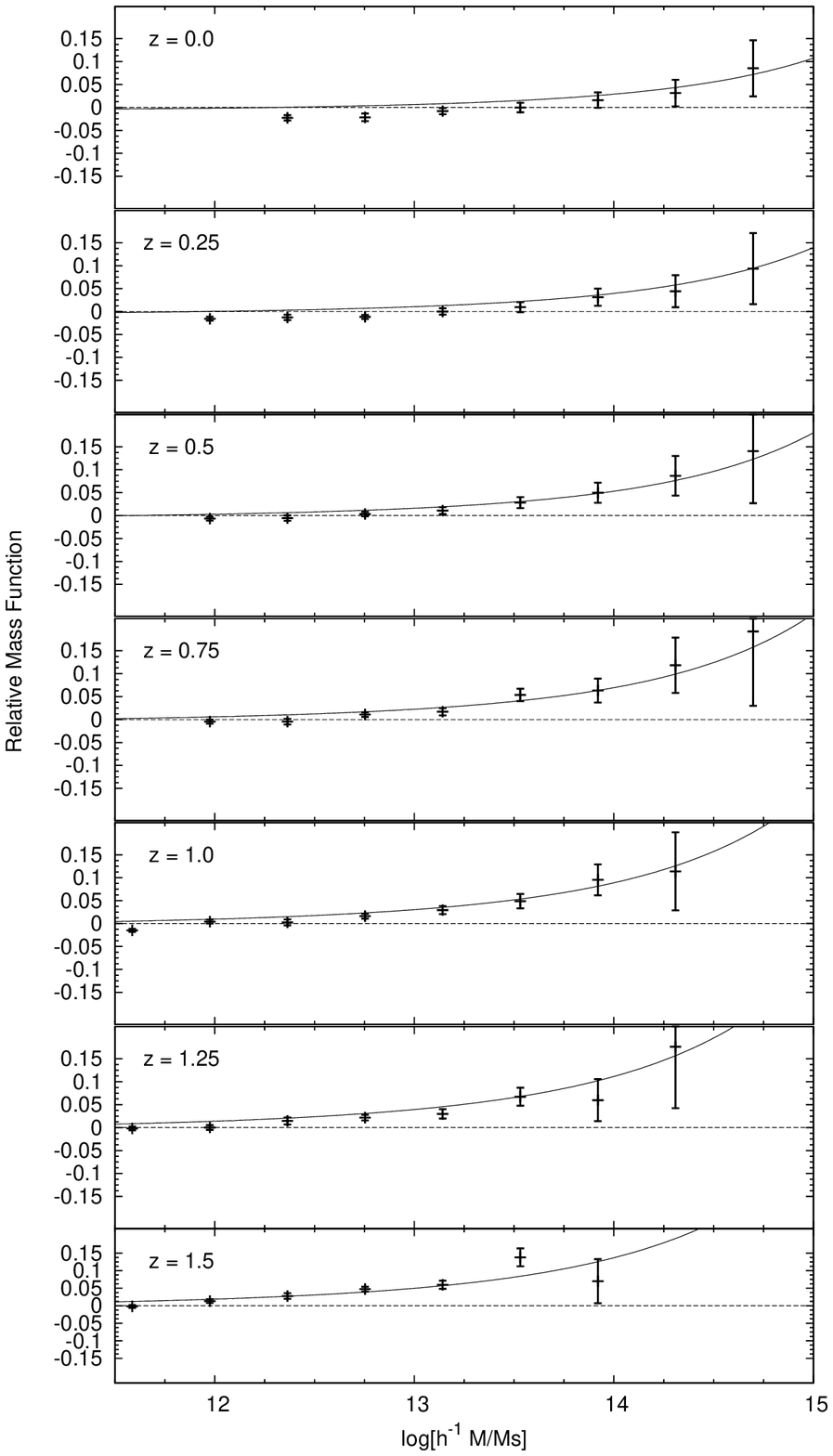,width=\columnwidth}}
  	\subfigure[Model B - $\Lambda$CDM]
		{\epsfig{figure=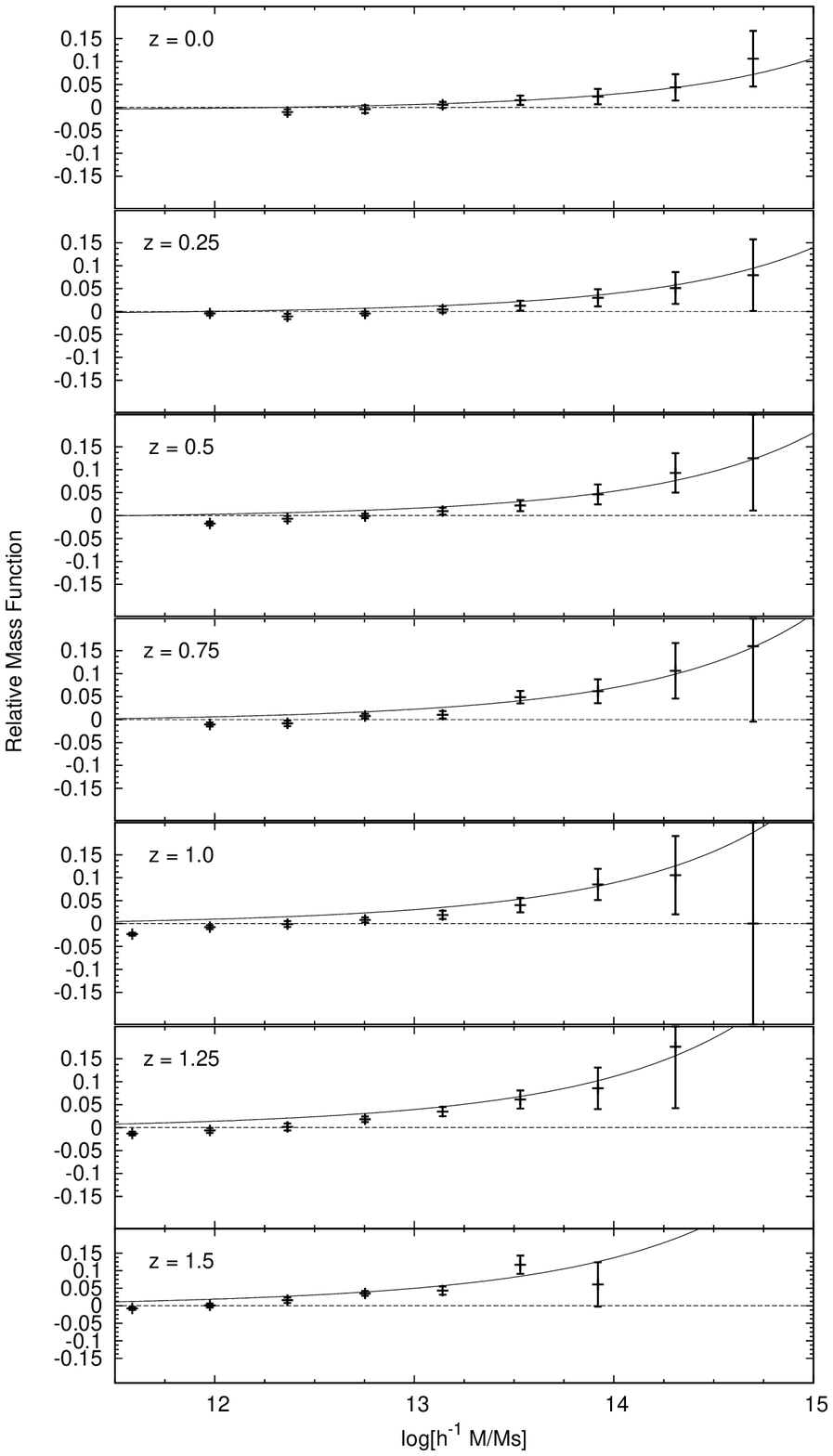,width=\columnwidth}}
        \caption{Relative mass functions for the power-law potentials.. 
                 The plots are labeled as in Table~\ref{tab:FandP}. 
                 Error bars are at one standard deviation and come 
		 from statistical counting uncertainties. The solid 
		 curve is the relative mass function 
		 of two \lcdm cosmologies with $\sigma_8=0.775$ 
		 and $\sigma_8=0.751$.}
\label{fig:massFunction}
\end{figure*}

Figure~\ref{fig:massFunction} shows the relative mass function from 
redshift $z=0$ to $z=1.5$ for the two models listed above. Both 
of these are compared to the $\Lambda$CDM simulation mass function. 
We will use the frequency density definition of the mass function,
\begin{equation}
F(M,z) \equiv \log{\frac{dn}{d \log{M}}},
\label{eq:massFunc}
\end{equation}
so that for model $i$ we may define the relative mass function as
\begin{equation}
RF(M,z)_i \equiv \left. \log{\frac{dn}{d \log{M}}} \right|_{i}
		- \left. \log{\frac{dn}{d \log{M}}} \right|_{\Lambda CDM}.
\label{eq:relativeMassFunction}
\end{equation}
We analyzed relative mass functions to reduce any systematic 
errors in the simulations, including those due to small box 
effects, such as missing tidal forces. We binned our distributions into 10
fixed logarithmic intervals from $10^{11.5}$ 
to $10^{15}$ $\hmsol$.
We only display values in bins for which we have complete data 
(i.e. the bin does not contain the minimum resolvable mass).
The uncertainties shown are obtained by summing in 
quadrature the 
individual statistical counting errors in the interacting and 
\lcdm cases. We see that at the present epoch, both models 
produce a greater number of the most massive halos while 
underproducing low-mass objects. At higher redshifts, both 
models produce greater numbers of all objects.
Since the relative mass function does not remain constant 
with redshift, it is distinguishable from a concordance 
cosmology with different fundamental parameters. 

\begin{figure}
	\epsfig{figure=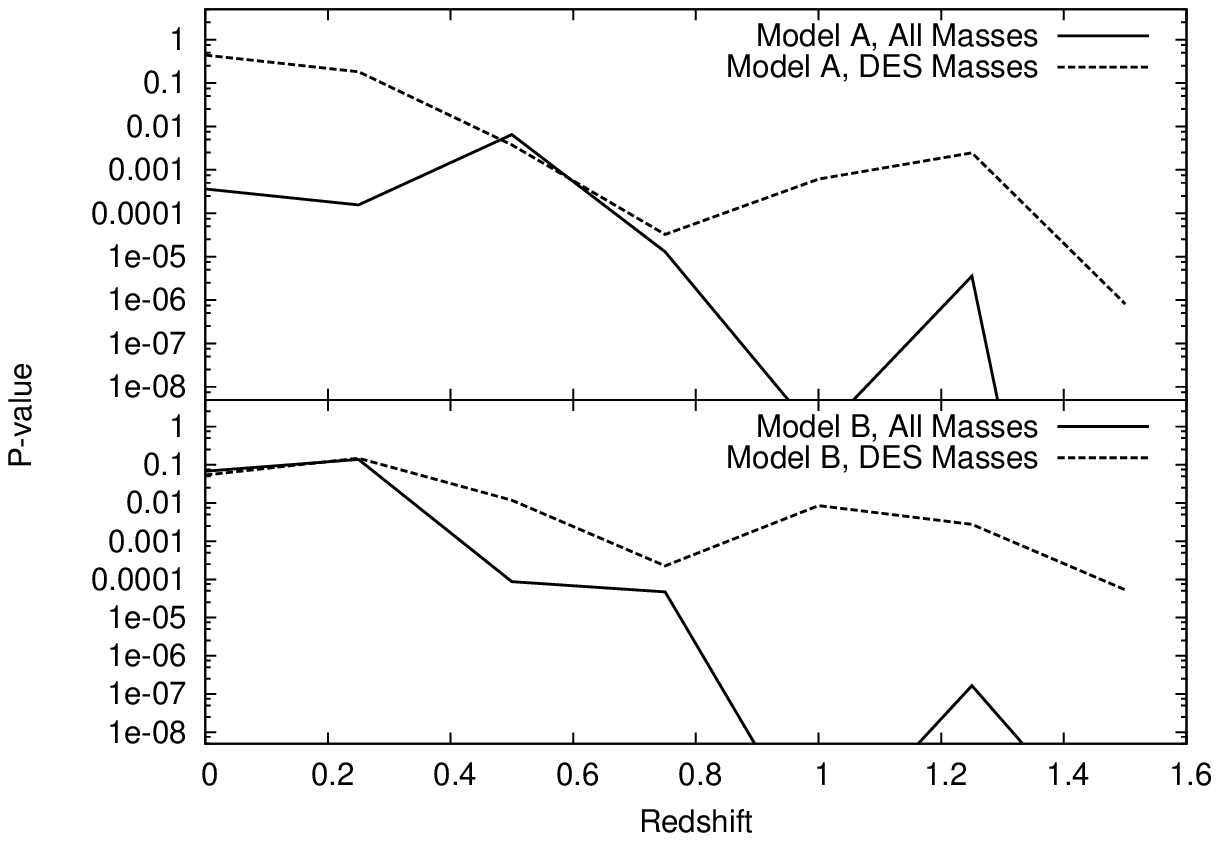, width=\columnwidth}
	\caption{Probability for the chi-square statistic 
		as a function of redshift for model A (top) and 
		model B (bottom) compared to $\Lambda$CDM. 
		The solid lines are from 
		including all resolvable masses, and the dashed lines 
		are from only considering objects 
		with $M_{FOF} > 10^{13.5} \hmsol$.}
\label{fig:pvalComp}
\end{figure}

To determine the significance of these mass function differences, 
we performed  
a $\chi^2$ likelihood test at each redshift.
For two indepedent frequency distributions $R$ and $S$,
\begin{equation}
\chi^2 = \sum_i \frac{(S_i-R_i)^2}{S_i+R_i},
\label{eq:chi-square}
\end{equation} 
where the sum takes place over all bins and the 
number of degrees of freedom is the total number 
of non-zero bins.
Figure~\ref{fig:pvalComp} shows the probability 
at each redshift that the 
frequency distributions from the interacting cases are consistent 
with the \lcdm case. We see that for almost all redshifts, the  
probability is exceedingly small, indicating that these 
coupled DM-DE models are well distinguished from \lcdm cosmology 
with this sample of objects.
As a function of redshift, the probability 
for both models generally decreases. 
For model $A$ at $z=0.5$, however, 
these distributions are not well separated, since this is 
the redshift at which the mass functions come closest.
Model $B$ remains indistinct from \lcdm until $z=0.5$, 
at which point the separation becomes progressively more evident.
The probabilities for both models drop below $10^{-10}$ at 
redshifts $z=0.75$ and $z=1.5$ and are not shown at higher redshifts.

To determine the feasibility of using a survey like DES to 
detect these models, we need to know the expected frequency 
distribution of observed objects in the survey, since the 
fundamental uncertainty arises from counting errors.
While the exact selection function for DES has not yet 
been determined, the survey is expected to capture $\sim10,000$ 
objects of mass $> 10^{13.5} \hmsol$ out to redshift $z\sim1.5$~\citep{DES}.
Our combined simulations produced roughly this many objects in this mass 
and redshift range.
Figure~\ref{fig:pvalComp} shows the $\chi^2$ 
probability when only considering mass bins above $10^{13.5} \hmsol$.
With this cut the probability suffers; however, 
we maintain the general trend of increased 
disparity with higher redshift. For both models, the 
greatest deviations occur at $z=0.75$ and $z=1.5$. The 
probabilities at these redshifts are below the 
common significance threshold and indicate that DES 
is capable of detecting these models. At lower redshifts, 
there is not a significant difference in the number of high-mass 
objects. Between these redshifts, 
the uncertainties in the largest mass bins are too large to make 
a confident distinction between the distributions.

It is possible to observe similar deviations in the 
mass function by changing the fundamental parameters 
of a \lcdm cosmology, such as $\Omega_{m0}$ 
and $\sigma_8$. We could not find any combination 
of fundamental parameters that reproduced these 
relative mass functions for all mass bins and redshifts. 
However, when examining masses within the DES limit, there 
are degeneracies. For example, Figure~\ref{fig:massFunction} 
shows the relative mass function, obtained using 
the~\citet{Warren} analytic mass function, of 
two \lcdm cosmologies with different values of $\sigma_8$. In this 
case, we compared a $\sigma_8$ of $0.751$, which we 
used in our simulations, to a cosmology with $\sigma_8=0.775$.
We chose this value to mimic the entire relative mass function at high 
redshift, but it does not capture the counts of lower-mass 
objects at low redshift. However, above $10^{13.5} \hmsol$, 
the uncertainties are large enough to permit this modified 
$\sigma_8$ to adequately fit the data. Thus, DES alone 
may not be able to distinguish between a universe with interacting DM-DE 
and a universe with higher $\sigma_8$.

\begin{figure}
	\epsfig{figure=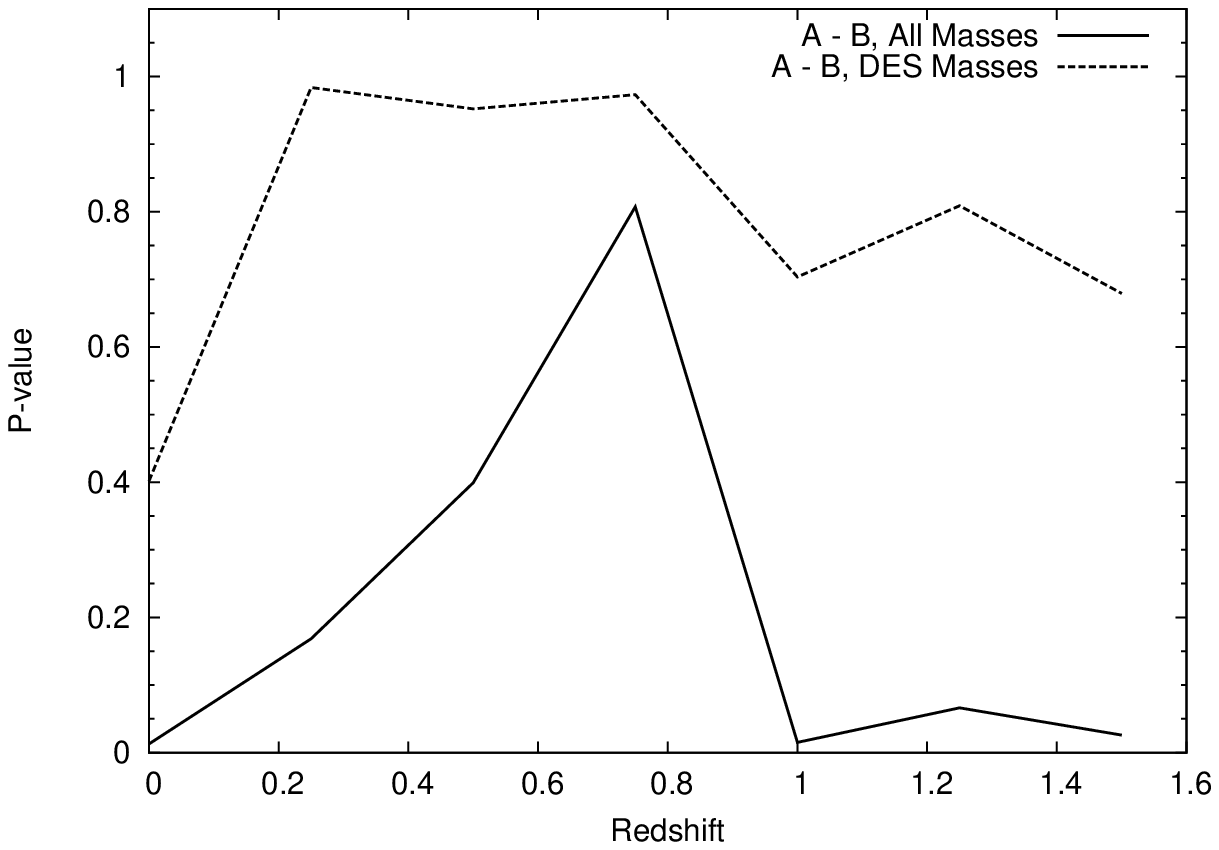, width=\columnwidth}
	\caption{Probability for the chi-square statistic 
		as a function of redshift for model A compared to 
		model B. 
		The solid line is from 
		including all resolvable masses, and the dashed line 
		is from only considering objects 
		with $M_{FOF} > 10^{13.5} \hmsol$.}
\label{fig:pvalDiff}
\end{figure}

While these two models are easily distinguishable from \lcdm cosmology, 
they are more difficult to differentiate from each other.
Figure~\ref{fig:pvalDiff} shows the $\chi^2$ probability when comparing 
these two models, both for all masses and DES-accessible masses.
When including all masses, the probabilities are significant
at redshifts $z=0.0$, $1.0$, and $1.5$. However, when examining 
only masses available to DES, the probabilities maintain a 
roughly constant performance and never reach a significant level.

\section{Conclusions}
\label{sec:Conclusion}
We have found that coupling dark matter to a dark energy scalar 
field produces significantly different mass functions at 
redshifts as high as $z=1.5$ relative to a \lcdm cosmology 
with the same set of fundamental parameters. 
This difference in the mass function follows from 
our analysis in SR08: an additional fifth force 
and a reduced particle Hubble drag lead to more 
structures than in \lcdm cosmologies at early times, and at late 
times will cause an overabundance of high-mass objects and 
a subsquent reduction in low-mass cluster counts.
By examining the mass function, we have developed a 
simple way of distinguishing these models. This 
analysis allows us to discover ways of 
further constraining different parameters of DM-DE coupling.

We have found that the statistical uncertainties in the mass 
function do not prevent the Dark Energy Survey from detecting 
this form of coupled dark matter and dark energy. 
Once the selection function for DES is known, 
a galaxy formation model can be applied 
and a more detailed study will need to take place. 
However, 
we have found that DES alone will have difficulty  
differentiating among different sets of 
parameters that control the coupling.

At high masses the statistical uncertainties 
may prevent DES from distinguishing between 
coupled DM-DE and \lcdm cosmologies with different values 
of $\sigma_8$. We can overcome this degeneracy in several ways. 
First, missions such as Planck~\citep{Planck} can 
independently constrain $\sigma_8$ and $\Omega_{m0}$. 
If DES prefers a higher value of $\sigma_8$ through 
the mass function, this may 
be explained by interacting dark matter and dark energy..
Secondly, more detailed measurements of the halo mass function 
from projects such as LSST~\citep{LSST} will tightly constrain the mass 
function at multiple redshifts. Also, DES itself may detect more 
clusters than our estimated 10,000. A universe with coupled DM-DE 
will then produce an apparently redshift-dependent $\sigma_8$.

It will also be necessary to compare these mass functions 
to those produced by modified General Relativity 
(such as those found in~\citet{Stabenau}), as 
both theories modify the Poisson equation, and hence 
can in principle have similar effects.

\section*{Acknowledgments}
The authors acknowledge support under a Presidential Early 
Career Award from the U.S. Department of Energy, 
Lawrence Livermore National Laboratory (contract B532720).
Additional support was provided by a DOE 
Computational Science Graduate Fellowship 
(DE-FG02-97ER25308) and the National Center for 
Supercomputing Applications.
The software used in this work was in part developed by the DOE-supported ASC 
/ Alliance Center for Astrophysical Thermonuclear Flashes at the University of Chicago.
\bibliography{ms}		
\bibliographystyle{apj}	
\nocite{*}

\end{document}